\documentclass[12pt,twoside,reqno]{article}
\usepackage[authoryear]{natbib}
\usepackage{graphicx}
\usepackage{amssymb,amsthm,amsmath,epsfig}
\usepackage{times}
\usepackage{bm}
\usepackage{color}
\usepackage{setspace}
\usepackage[utf8]{inputenc}
\usepackage{lscape}

\usepackage[nottoc]{tocbibind}

\usepackage[english]{babel}

\usepackage{actuarialangle}
\usepackage{actuarialsymbol}

\newcommand\x{3}
\usepackage[top = \x cm, bottom = \x cm, left = \x cm,right = \x cm]{geometry}

\newtheorem*{rel*}{Relationship}
\newtheorem*{genrel*}{Generalized Relationship}

\usepackage{authblk}

\usepackage{fancyhdr}

\usepackage{changes}
\usepackage{lipsum}
\definechangesauthor[name={Trifon I. Missov}, color=orange]{TIM}

\title{Makeham Mortality Models as Mixtures}

\author[1]{Silvio C. Patricio\thanks{silca@sam.sdu.dk}}
\author[1]{Trifon I. Missov}
\affil[1]{\small{Interdisciplinary Center on Population Dynamics, University of Southern Denmark}}

\date{}

\begin{document}
\maketitle
    
\begin{abstract}
\noindent \noindent\textbf{Background:}
The Makeham term serves as a fundamental component in mortality modeling, offering a constant additive hazard that accounts for background mortality factors usually unrelated to the aging process. This term, widely employed in mortality analysis, provides a crucial mechanism for capturing mortality risks unrelated to age-related deterioration.
\textbf{Objective:}
The objective of this paper is to explore the relationship between Makeham mortality models and competing risk frameworks. It aims to investigate how Makeham models, which are widely used for studying mortality, can be understood and analyzed within the context of competing risks. The paper seeks to provide insights into the mathematical properties, interpretation, and applicability of Makeham models in modeling mortality risks associated with various causes of death. Additionally, the paper aims to demonstrate formally that competing risk models can be represented as mixture models, thereby facilitating a deeper understanding of mortality dynamics. 
\textbf{Contribution:}Makeham mortality models, when represented as mixtures, have a straightforward specification that can easily be extended to account for unobserved heterogeneity. They provide a semantically and computationally convenient platform to disentangle senescent from extrinsic (premature) mortality. By expressing Makeham models as a convex combination of probability distributions, we are able to estimate the age profile of premature mortality, especially at the oldest ones, where we intuitively assume that most deaths are senescent. We are also able to estimate the senescent mortality component, which is the one to focus on when studying the aging process.  
\end{abstract}

\section{Introduction}\label{Introduction} 
Death can occur due to various causes, each characterized by its own failure time. Individuals are exposed to a finite number of such causes, and the interplay of the latter determines the observed length of life.

Let there be a finite number of causes of death labeled $c_1, c_2, \dots, c_n.$ Suppose a non-negative random variable $T_k$ with a hazard function $h_k(x)$ captures the age at death from cause $c_k$, $k=1, 2, \dots, n$, when $c_k$ is the only cause of death. Failure time $T_k$ measures the `'absolute potency'' of cause $c_k$. The length of life for an individual is captured then by the random variable $Y = \min(T_1, \dots, T_n)$, whose hazard function, $\mu(x)$, reflects the competing $n$ cause-specific risks of dying \citep[see, for example,][]{chiang1961probability,berman1963note,gail1975review}. 

Makeham models are widely used for studying mortality as they account for a simple and straightforward competing-risk framework. They are characterized by a hazard function of the type
\begin{equation} \label{hazard-general}
\mu(x) = h(x) + c \,,
\end{equation}
where $h(x)$ is another hazard function. In the context of mortality, $h(x)$ most often pertains to the aging process, while $c$, the Makeham term, accounts for extrinsic (or \textit{background}) hazards that usually capture premature deaths. When

\begin{equation}
h(x) = ae^{bx} \,,
\end{equation}
(\ref{hazard-general}) reduces to the Gompertz-Makeham (GM) model \citep{gompertz1825xxiv,makeham1860law}, while for

\begin{equation}
h(x) = \frac{ae^{bx}}{1+\frac{a\gamma}{b}(e^{bx}-1)} \,,
\end{equation}
(\ref{hazard-general}) yields the gamma-Gompertz-Makeham ($\Gamma$GM) model \citep{VauManSta79}. \citet{Bea59} and \citet{Kan94} provide alternative logistic curves with different asymptotics. The hazard $h(x)$ can be extended by an additive component that reflects infant and childhood mortality, as in Siler's model \citep{siler1979}

\begin{equation}
h(x) = a_1e^{-b_1x} + ae^{bx} \,.
\end{equation}

The GM, $\Gamma$GM, and Siler models reflect a competing risk framework: an individual dies either as a result of biological processes at early or late ages or due to some extrinsic risk $c$, whatever strikes first. 

The Makeham term in all models described by the general framework (\ref{hazard-general}) accounts for a competing exponentially distributed risk. \cite{elandt1976conditional,gail1975review,hakulinen1977example} discuss mathematical properties and practical implications of assuming independence among failure times, as well as the consequences when this assumption is violated. They do not study, though, the representation of competing risk models as mixtures, in which every distribution characterizes the distribution of deaths due to a specific out of of $n$ potential competing risks.


\section{The Relationship} \label{sec_relationship}


Suppose a population is subject to $n$ causes of death $c_1, c_2, \dots, c_n$, where each individual is characterized by a corresponding vector random variable $\bm T = (T_1, T_2, \dots, T_n)'$ representing the hypothetical failure time from each cause of death in the absence of other causes. Then, a non-negative continuous random variable $Y$, capturing the length of life for an individual, can be represented as  $Y = \min(T_1, T_2, \dots, T_n)$, its hazard function can be expressed as
\begin{equation} \label{hazard-gen}
    \mu(x) = h(x, 1)+h(x, 2)+\cdots+h(x, n) \,,
\end{equation}
and its probability density function (PDF) can be represented as
\begin{equation} \label{fx-gen}
    f(x) = \sum_{j=1}^n h(x, j) \exp \left\{ -\sum_{k=1}^n H(x, k) \right\}\,,
\end{equation}
where $H(x, k) = \int\limits_{0}^{x} h(t, k)dt$ is the $k$-th cumulative sub-hazard function. (\ref{fx-gen}) implies that the PDF of $Y$ can be expressed as a mixture

\begin{equation} \label{fx-gen2}
    f(x) = \sum_{j=1}^n \pi_j g_j(x) \,,
\end{equation}
where $g_j(x) = f(x \, | \, j)$ denotes the PDF of the lifespan for individuals who die from $c_j$, $j=1,2, \dots, n$ while being exposed to all $n$ risks, and $\pi_j$ is the probability that cause $c_j$ strikes first among all competing causes.


\section{Proof}\label{Proof} 

We want to demonstrate that a competing risk model can be represented as a mixture model. In a competing risk setting, individuals are exposed to multiple causes of death, and death occurs when the first cause strikes. This is analogous to a mixture model, where each cause of death corresponds to a component distribution, and lifetimes are modeled as a mixture of these component distributions.

We will prove that the density of the lifetime distribution, representing the time until the first from multiple competing causes strikes, can be expressed as a convex combination of probability distributions corresponding to dying from each cause of failure in the presence of the risks from all others. Each component distribution corresponds to the lifetime of individuals who die from a specific cause. The weights $\pi_j$ in the convex combination capture the probabilities of each cause $c_j$ striking first.

\bigskip

\begin{proof}
    It is given that a population is subjected to $n$ causes of death, labeled $c_1, c_2, \dots, c_n$, and each individual is characterized by a vector of random variables $\mathbf{T} = (T_1, \dots, T_n)'$, representing the failure times $T_j$ according to each cause of death $c_j$ in the absence of other competing causes, $j = 1, \ldots, n$. We denote the hazard function for each $T_j$ by

    \begin{equation*}
        h_j(t) = \lim_{\Delta \downarrow 0} \frac{1}{\Delta} \mathbb{P}(t < T_j \leq t + \Delta | T_j > t)\,,
    \end{equation*}
    and the corresponding survival function by $S_j(t)$. The survival function of $\mathbf{T}$ is given by 
    
    \begin{equation*}
        S_{\mathbf{T}}(t_1, \cdots, t_n) = \mathbb{P}\left( \bigcap_{j=1}^n [T_j > t_j] \right),
    \end{equation*}
    which satisfies $S_{\mathbf{T}}(0, \cdots, 0) = 1$ and $S_{\mathbf{T}}(\infty, \cdots, \infty) = 0$, with each $t_j \in (0, \infty)$. The function $S(\cdot)$ is continuous from the right and monotonically non-increasing in each argument.

    Note that we cannot observe the failure times $T_1, \dots, T_n$ simultaneously. As a result, $S_{\mathbf{T}}(t_1, \cdots, t_n)$ cannot be observed, nor can its form be tested \citep{elandt1976conditional}. Instead, as highlighted by \cite{moeschberger1971life}, what we observe is the individual lifetime denoted by $Y = \min(T_1, \dots, T_n)$. The corresponding survival function is given by
    \begin{equation} \label{eq:proof_surv}
        S_{Y}(t) = \mathbb{P}\left( Y > t \right) = S_{\mathbf{T}}(t, \cdots, t) \,,
    \end{equation}
    where $t$ is the observed failure time.
    
    With the assumption that $\mathbb{P}(T_i = T_j) = 0$ for all $i \neq j$, we can define the random variable $J$ as the index of the smallest $T_i$. Therefore, $J = j$ implies that cause $c_j$ is responsible for the death, and $T_j < T_i$ for all $i \neq j$, $j = 1, \dots, n$. By applying the law of total probability and Bayes' theorem to (\ref{eq:proof_surv}), we obtain
    \begin{equation*}
        S_{Y}(t) = \sum_{j = 1}^n \mathbb{P}\left( Y > t, J = j\right) = \sum_{j = 1}^n  \mathbb{P}\left( Y > t \, | \, J = j\right)\mathbb{P}\left(J = j\right).
    \end{equation*}
    The function
    
    \begin{equation*}
        S(t \, | \, j) = \mathbb{P}\left( Y > t \, | \, J = j\right) = \frac{\mathbb{P} (t < T_j < \min_{i \neq j} T_i)}{\mathbb{P} (T_j < \min_{i \neq j} T_i)},
    \end{equation*}
    represents the cause-specific survival function for cause $c_j$, while the quantity 
    
        \begin{equation*}
		\pi_j = \mathbb{P}\left(J = j\right) = \mathbb{P} (T_j < \min_{i \neq j} T_i)
	\end{equation*}
	is the probability that cause $c_j$ strikes first among all competing causes. Thus, 
    
    \begin{equation} \label{SY}
    S_{Y}(t) = \sum_{j = 1}^n \pi_j S(t \, | \, j) \,.
    \end{equation}
    Assuming $S_{\mathbf{T}}$ is differentiable, we can express the density of $Y$ as
    
    \begin{equation} \label{fY}
        f_Y (t) = - \frac{d}{dt}S_{Y}(t) = \sum_{j = 1}^n f(t, j) = \sum_{j = 1}^n \pi_j f(t \, | \, j) \,,
    \end{equation}
    where $f(t \, | \, j) = - \frac{d}{dt} S(t \, | \, j) $ represents the density of lifetimes due to competing cause $c_j$. Integrating  $f(t, j)$ with respect to $t$, we can derive $\pi_j$. The hazard function of $Y$ is given by
    
    \begin{equation} \label{hY}
        h_Y(t) = \frac{f_Y (t)}{S_Y(t)} = \sum_{j = 1}^n h(t,j) \,,
    \end{equation}
    where $h(t, j) = \frac{f(t,j)}{S_Y(t)}$. This function is sometimes called the $j$th sub-hazard rate and should not be confused with the function $h(t \, | \, j) = - \frac{d}{dt} \log S(t \, | \, j) = f(t \, | \, j)/S(t \, | \, j)$.
    
    (\ref{SY})-(\ref{hY}) illustrate that, no matter if the failure times $T_1, \dots, T_n$ are independent or not, the distribution of lifetimes can be expressed as a convex combination of $n$ probability distributions, each representing the time to death for individuals affected by a given cause $c_j$, $j=1,\ldots,n$.
    
\end{proof}

\subsection*{Note 1: The assumption on the identity of the forces of mortality}
As previously mentioned, the function $ S_{\mathbf{T}}(t_1, \cdots, t_n) $ is not directly observable, nor can its form be tested. In practical applications, an additional assumption on the identity of the forces of mortality is typically made:
\begin{equation*}
    h(t, j) = h_j(t), \quad j = 1, \dots, n.
\end{equation*}
This assumption implies that $ S_Y(t) $ can be expressed as a product of the individual survival functions $ S_j(t) $:
\begin{equation*}
    S_Y(t) = S_{\mathbf{T}}(t, \cdots, t) = \prod_{j = 1} ^ n S_j(t) \,.
\end{equation*}
This assumption suggests that the events $ [T_j > t], \, j = 1, \dots, n $ are independent. Note that this does not necessarily imply independence of the random variables $ T_1, \dots, T_n $ \citep[see, for example, ][]{gail1975review}.

As noted by \cite{elandt1976conditional}, directly observing the 'time-to-death from a given cause of death' random variable is not feasible, and estimating its associated survival function requires additional assumptions. In addition,  \cite{crowder1991identifiability} demonstrated that even when the marginal survival functions $S_j(t_j)$ of the failure time $T_j$ are known, observations of $Y$ and $J$ do not determine their joint survival function $S_{\bm T}$. Therefore, assuming independence between events like $ [T_j > t], \, j = 1, \dots, n $, or even between the random variables $T_i$ and $T_j$, for $i \neq j, \, j = 1, \dots, n$, though not necessarily reflecting the true interaction mechanism among different causes of death, provides a reasonable framework for distinguishing between extrinsic and senescent mortality.

\subsection*{Note 2: On independence assumption for cause-specific failure times}

When we assume that the random variables $T_j, j = 1, 2, \dots, n$, are mutually independent, the survival function $S_{\bm T}(t_1, \cdots, t_n)$ simplifies to $\prod_{j=1}^n S_j(t_j)$, making the proof of the relationship rather straightforward. However, assuming independence of cause-specific failure times has at least two implications: first, the associated cause-specific risks of dying act independently of one another, and second, eliminating one cause of death does not affect the force of mortality of the other cause. However, the adequacy of this assumption varies depending on the specific cause of death and its complex relationship with other causes. A comprehensive model should acknowledge that (a) a simple constant term, such as the Makeham term, may not accurately capture the complex mechanisms underlying premature deaths; (b) lifetimes may not be independent, highlighting the need for a more sophisticated modeling approach; and (c) the dependence of lifetimes, reflecting behavioral, social, and environmental risk factors, may vary across different age groups.

According to \cite{cox1959analysis}, when considering two risks, data collected on variables $Y$ and $J$ will not provide evidence contradicting the assumption that the failure times associated with these risks are independent of each other. In other words, observations of $Y$ and $J$ will not suggest dependence between the failure times linked to these risks.

Even when circumventing the assumption of independence of lifetimes by using the method proposed by \cite{gumbel1967some} for random variables $T_1$ and $T_2$ following extreme value distributions, the estimation of correlation parameters is complex \citep{tawn1988bivariate}. Thus, applying such a model to aggregate (life-table) mortality data presents significant challenges.

\section{Related results}\label{related_results} 
Makeham models are a special case of (\ref{hazard-gen}) when $n=2$, $h_1(x) = c$ and $h_2(x) = h(x)$.
For the Makeham models, we consider a non-negative continuous random variable $Y$ characterized by a Makeham hazard function as described in equation (\ref{hazard-general}). This hazard function represents the combined risk of two causes of death: extrinsic, denoted by $c_1$, and biological (senescent), denoted by $c_2$. Each cause is described by its own failure time, captured by the not necessarily independent random variables $T_1$ and $T_2$, respectively. Subsequently, the probability density function (PDF) of $Y = \min(T_1, T_2)$ can be represented as:

\begin{equation} \label{eq:fx-gen_mak}
    f(x) = c\exp\left\{ -cx - H(x) \right\} + h(x) \exp\left\{ -cx - H(x) \right\},
\end{equation}
where $H(x) = \int\limits_{0}^{x} h(t)dt$ is the cumulative hazard function of $T_2$. From (\ref{eq:fx-gen_mak}), it is easy to represent a Makeham model of type (\ref{hazard-general}) as a mixture model:
\begin{equation} \label{fx-gen2_mak}
    f(x) = \pi g_1(x) + (1-\pi)g_2(x),
\end{equation}
where $g_j(x) = f(x \, | \, j)$ denotes the PDF of the lifespan for individuals who die from $c_j$, $j=1,2$, being exposed to both the senescent and the premature risk of dying. 

Each component of the mixture (\ref{fx-gen2_mak}) has a straightforward interpretation. For example, 
\begin{equation}
    \pi = \int_0^\infty c\exp\left\{ -cx - H(x) \right\} dx\,, \label{eq_pi_general}
\end{equation}
known as the mixing proportion \citep{mclachlan2019finite}, represents the probability of dying from an extrinsic cause in the presence of a competing senescent risk of dying, i.e., $\pi$ is the \textit{premature mortality prevalence}. On the other hand, the functions
\begin{equation} \label{g1g2}
    g_1(x) = \frac{c}{\pi} \exp\left\{ -cx - H(x) \right\} \hspace{3mm} \text{and} \hspace{3mm} g_2(x) = \frac{h(x)}{1-\pi} \exp\left\{ -cx - H(x) \right\}
\end{equation}
capture the PDFs of the distribution of deaths according to each general cause: in the presence of competing premature and senescent risks of dying, $g_1$ represents the PDF of non-aging-related (premature) deaths, while $g_2$ denotes the PDF of aging-related (senescent) deaths. 

By representing the class of models (\ref{hazard-general}) as a mixture of two probability distribution functions, we can determine the threshold age, $x^*$, that separates the age interval with prevailing premature deaths from the one with predominant senescent mortality: 

\begin{equation}
x^* = \max\{ \max\{ x: h(x) = c \}, 0 \}\,.  
\end{equation}
We can also express the proportion of non-aging-related deaths at age $x$ as 

\begin{equation}
p(x) = c/(h(x)+c)\,.
\end{equation}
$p(x)$ reflects the age-specific prevalence of premature mortality. Table \ref{tab:models} presents closed-form expressions for $x^*$ and $p(x)$ for four popular Makeham mortality models.
 
\begin{table}[!htp]
\centering
{\small
\caption{Expressions for the threshold age, $x^*$, separating the age intervals with prevailing premature and senescent mortality, as well as the proportion of non-aging-related deaths at age $x$, $p(x)$, in Gompertz-Makeham, gamma-Gompertz-Makeham, Beard-Makeham, Kannisto-Makeham, and Siler model settings.}\label{tab:models}
\begin{tabular}{lccc}   \hline
Model  & $h(x)$ & $x^*$ & $p(x)$ \\ \hline
Gompertz-Makeham &  $ae^{bx}$ & $\frac{1}{b} \ln \frac{c}{a}$ & $\frac{c}{ae^{bx}+c}$\\
$\Gamma$-Gompertz-Makeham &  $\frac{ae^{bx}}{1+\gamma \frac{a}{b}\left( e^{bx}-1\right)}$ & $\frac{1}{b} \ln \frac{c(b-a\gamma)}{a(b-c\gamma)}$ & $\frac{c}{\frac{ae^{bx}}{1+\gamma \frac{a}{b}\left( e^{bx}-1\right)}+c}$ \\
Beard-Makeham & $\frac{ae^{bx}}{1+kae^{bx}}$ & $\frac{1}{b}\ln \frac{c}{a(1-kc)} $ & $\frac{c}{\frac{ae^{bx}}{1+kae^{bx}}+c}$ \\
Kannisto-Makeham & $\frac{ae^{bx}}{1+ae^{bx}}$ &  $\frac{1}{b}\ln \frac{c}{a(1-c)}$ & $\frac{c}{\frac{ae^{bx}}{1+ae^{bx}}+c}$ \\
Siler & $a_1e^{-b_1x} + a_2e^{b_2x}$ & no closed-form & $\frac{c}{a_1e^{-b_1x} + a_2e^{b_2x}+c}$ \\
\hline
\end{tabular}    }
\end{table}

Note that assuming a constant extrinsic risk of dying, does not imply a constant force of mortality for the distribution of premature deaths (in the presence of a competing senescent risk). The force of mortality for this subpopulation is given by
\begin{equation}
    h(x \, | \, 1)= \frac{\exp\left\{ -cx - H(x) \right\}}{\int_x^\infty  \exp\left\{ - cy - H(x) \right\} dy}\,.
\end{equation}

In many Makeham mortality models, we can derive closed-form expressions for different characteristics of the distributions of deaths in the mixture. For example, the remaining life expectancy at age $x$ is given by 

\begin{equation} \label{remle}
e_x = \frac{1}{S(x)}\int_0^\infty S(x+t) \, dt \,, 
\end{equation}
where $S(\cdot)$ is the survival function. Substituting $S(x) = S(x \, | \, 1) = \int_x^\infty g_1(y) \, dy$ and $S(x) = S(x \, | \, 2) = \int_x^\infty g_2(y) \, dy$ in (\ref{remle}) yields $e_x$ for the subpopulation struck by the premature and senescent competing risk, respectively. The modal age at death for these subpopulations corresponds to the maximum of $g_1(x)$ and $g_2(x)$, respectively. Note that for any non-decreasing $h(x)$, the modal age at death for the non-senescent subpopulation is zero. For the senescent subpopulation (with PDF $g_2$), the modal age at death is the age where the overall force of mortality equals the relative derivative of $h(x)$ with respect to age: 

\begin{equation}
h(x)+c = \frac{dh(x)/dx}{h(x)}\,. 
\end{equation}
For the Gompertz-Makeham model, if $b>c$, the senescent modal age at death is given by 

\begin{equation}
M_{GM} = \frac{1}{b} \, \ln \frac{b-c}{a}\,, 
\end{equation}
while for the gamma-Gompertz-Makeham model, it is given by 

\begin{equation}
M_{\Gamma GM} = \frac{1}{b} \, \ln \left( \frac{b}{a} \cdot \frac{b-a\gamma - c \left(1 - \frac{a\gamma}{b}\right)}{b+a\gamma c}\right)\,.
\end{equation}
When $c = 0$, the expressions for $M_{GM}$ and $M_{\Gamma GM}$ reduce to the modal age at death for the Gompertz and the gamma-Gompertz model, respectively (see \cite{missov2015gompertz} for details). The modal age at death can also be expressed for the Beard-Makeham model as 

\begin{equation}
M_{BM} = \frac{1}{b} \, \ln \frac{b-c}{a(1+ck)}\,, 
\end{equation}
while in the Siler model setting, there is no closed-form expression. Note that, given the improvements in human mortality and the rise of longevity, we may observe a decreasing trend for $c$ over time. This leads to a convergence of the overall modal age at death to the senescent one.

We can also derive closed-form expressions for $\pi$ in a Gompertz-Makeham and a gamma-Gompertz-Makeham setting. From (\ref{eq_pi_general}), we can express $\pi = c e_0$, where $e_0$ is the life expectancy at birth of the Makeham model. Using the life expectancy formulae in \cite{castellares2020closed}, we can derive expressions for $\pi$ in terms of special functions. For the Gompertz-Makeham model, we have that

\begin{equation} \label{eq:pi_makeham}
    \pi = \frac{c}{a} \left( \frac{a}{b}\right)^{\frac{c}{b}} e^\frac{a}{b} \Gamma \left( - \frac{c}{b}, \frac{a}{b} \right)
\end{equation}
where $\Gamma(u, x) = \int_x^\infty t^{u-1}e^{-t}dx$, $x>0$, $u \in \mathbb{R}$ is a complementary incomplete gamma function. For the gamma-Gompertz-Makeham model, the closed-form expression for $\pi$ is given by
\begin{equation}
    \pi = \frac{c \gamma}{b+c \gamma} \, {}_2F_1 \left( \frac{1}{\gamma}, 1; \frac{1}{\gamma}+\frac{c}{b} + 1; 1-\frac{a\gamma}{b}\right)
\end{equation}
where ${}_2F_1(m,p;q; z)= \frac{\Gamma(q)}{\Gamma(p) \Gamma(q-p)} \int^{1}_{0} u^{p-1} (1-u)^{q-p-1} (1-zu)^{-m} du$ is the Gaussian hypergeometric function.

\subsection{The Gompertz-Makeham model}

The Makeham term has first been used to adjust Gompertz mortality estimates \citep{makeham1860law}, resulting in the Gompertz-Makeham model. A non-negative continuous random variable $Y$ has a Gompertz-Makeham distribution if its hazard function is given by
\begin{equation} \label{eq:hazard_makeham}
    \mu(x) = a e^{b x} + c, \quad a, b>0, \hspace{.2cm} c \geq 0; \hspace{.4cm} x \geq 0,
\end{equation}
and survival function
\begin{equation} \label{eq:survival_makeham}
    S(x) = \exp \left\{ -\frac{a}{b} \left( e^{b x } -1 \right) - c x \right\} \,.
\end{equation}

\begin{figure}[htb!]
    \centering
    \includegraphics[width=0.9\textwidth]{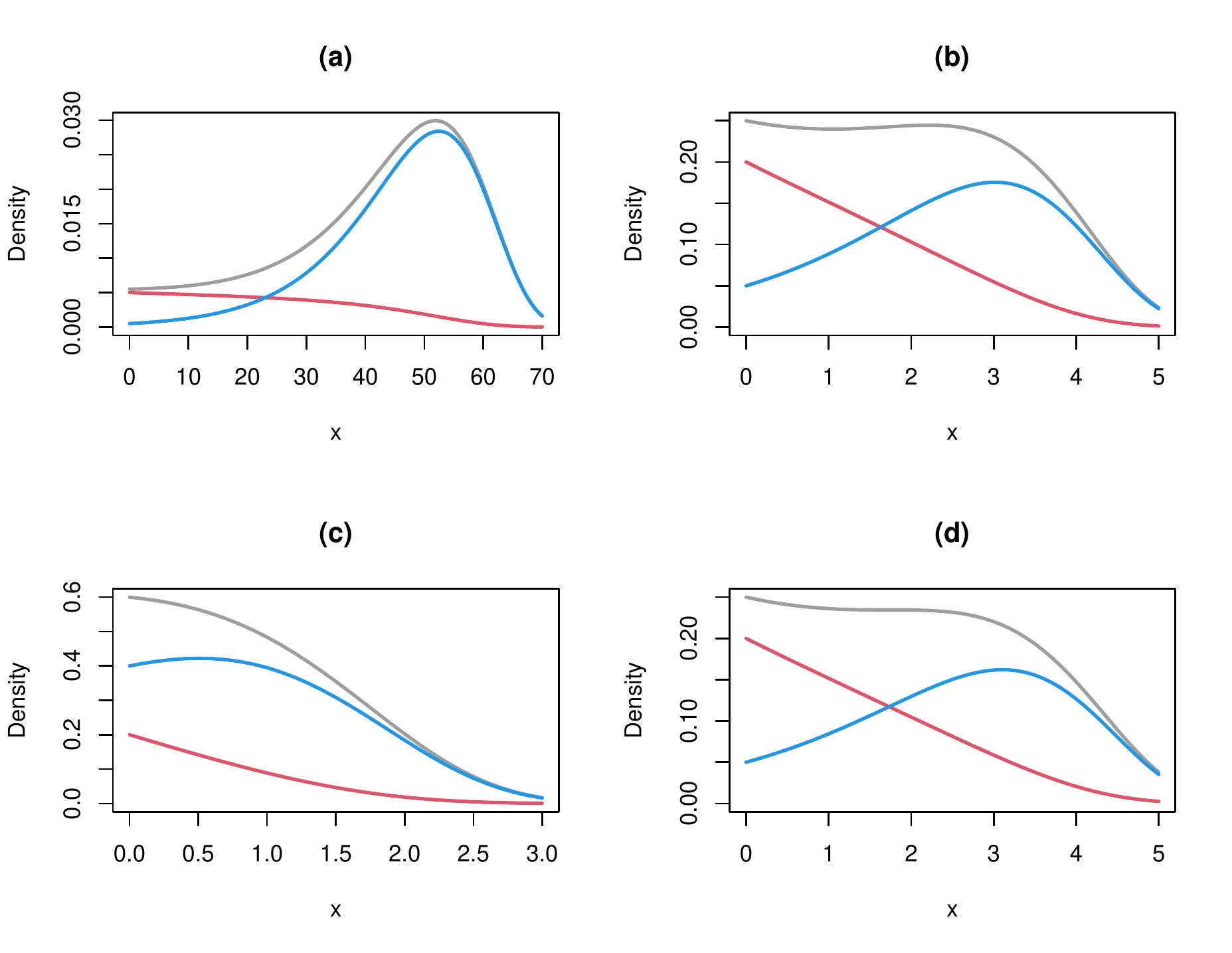}
    \caption{PDF of the Gompertz-Makeham distribution as a mixture. The grey line presents $f$; the red line indicates $g_1$, and the blue line represents $g_2$. For (a) we have $a = 0.0005$, $b=0.1$ and $c = 0.005$, for (b) we have $a = 0.05$, $b=0.85$ and $c = 0.2$, for (c) we have $a = 0.4$, $b=0.8$ and $c = 0.2$, and for (d) we have $a = 0.05$, $b=0.8$ and $c = 0.2$. }
    \label{fig:mix}
\end{figure}

From (\ref{eq:hazard_makeham}) and (\ref{eq:survival_makeham}), it is straightforward to derive the Gompertz-Makeham PDF: 
\begin{equation} \label{eq:pdf_makeham}
    f(x) =  c\exp \left\{ -\frac{a}{b} \left( e^{b x } -1 \right) - c x \right\}+ a e^{b x}\exp \left\{ -\frac{a}{b} \left( e^{b x } -1 \right) - c x \right\}.
\end{equation}
According to \cite{castellares2022gompertz}, there are four different shapes for $f(x)$: (a) it may have a local maximum that is not located at the boundary; (b) it may have a local maximum both at the boundary and inside the parameter space; (c) it may have a global maximum at the boundary; and (d) it may have a global maximum at the boundary and an inflection point inside the parameter space \citep{norstrom1997gompertz}. The four shapes are displayed in Figure \ref{fig:mix}. While only panel (a) in Figure \ref{fig:mix} corresponds to the pattern observed in human mortality, panels (b)-(d) were included to illustrate the diverse dynamics of mortality when decomposing senescence and premature mortality.


{}

The distribution of a Gompertz-Makeham random variable $Y$ can be perceived as a mixture, i.e., its PDF (\ref{eq:pdf_makeham}) can be represented as a convex combination of two probability distributions
\begin{equation}
    f(x) = \pi g_1(x) + (1-\pi) g_2(x),
\end{equation}
where

\begin{equation} \label{gompmake_g1g2}
    g_1(x) = \frac{c}{\pi}\exp \left\{ -\frac{a}{b} \left( e^{b x } -1 \right) - c x \right\}, \quad g_2(x) = \frac{a e^{- (c-b) x}}{1-\pi}\exp \left\{ -\frac{a}{b} \left( e^{b x } -1 \right) \right\}
\end{equation}
and $\pi$ is given by (\ref{eq:pi_makeham}).

Under the assumption of identity of the forces of mortality, the survival function takes the form
\begin{equation}
    S(x) = S_{1}(x) \, S_{2}(x) = \pi G_1(x) + (1-\pi) G_2(x)
\end{equation}
with $G_j(x) = S(x \, | \, j)= \int\limits_x ^ \infty g_j (t) dt$ and $S_j(x) = \exp\left\{-\int_0^x h_j(t) dt\right\}$, for $j = 1,2$.


Figure \ref{fig:mix} shows the decomposition of the Gompertz-Makeham PDF into senescent ($g_2$ density) and non-senescent ($g_1$ density) components, from which we can see how non-senescent deaths shift the mode of the distribution (also known as \textit{modal age at death}) to the left, sometimes even leading to a second peak of the distribution (panel (b) in Figure \ref{fig:mix}). Note that although the competing risks are Gompertz and Exponential, (\ref{gompmake_g1g2}) suggests that the Gompertz-Makeham density is not represented by a convex combination of a Gompertz and an Exponential distribution.

\begin{figure}[htb!]
    \centering
    \includegraphics[width=0.9\textwidth]{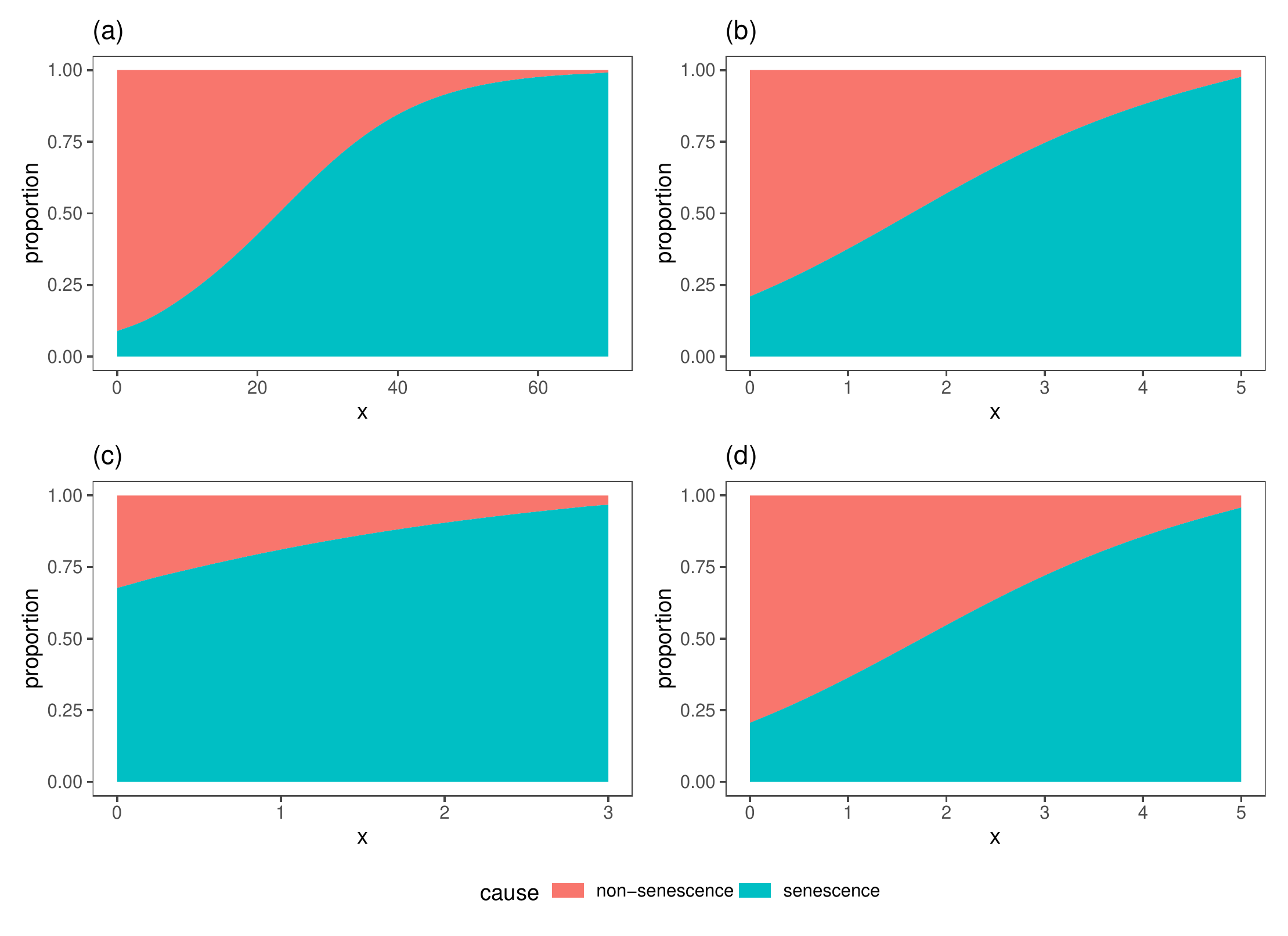}
    \caption{Proportion of senescent and non-senescent deaths by age in the Gompertz-Makeham setting. For (a) we have $a = 0.0005$, $b=0.1$ and $c = 0.005$, for (b) we have $a = 0.05$, $b=0.85$ and $c = 0.2$, for (c) we have $a = 0.4$, $b=0.8$ and $c = 0.2$, and for (d) we have $a = 0.05$, $b=0.8$ and $c = 0.2$.}
    \label{fig:prop}
\end{figure}

By representing the Gompertz-Makeham as a mixture of two distributions, we are able to quantify the overall proportion of non-senescent deaths, given by the quantity $\pi$, and also the proportion of non-senescent deaths at age $x$, equal to $p(x)$. For the Makeham-Gompertz model, the age-specific proportion of premature deaths can be expressed as $p(x) = c/\left( ae^{bx}+c \right)$. Figure \ref{fig:prop} shows the proportion of senescent and non-senescent deaths by age in a Gompertz-Makeham model for different parameter combinations. 

{}

The function $p(x)$ aids in estimating the threshold age, $x^*$ that separates the ages with predominant non-senescent deaths from the ages with predominant senescent deaths. For the Gompertz-Makeham model, 

\begin{equation}
x^*  = \left\{
\begin{array}{ l l }
    \frac{1}{b}\ln\frac{c}{a} & c \geq a \\
    0 & c < a \\
    \end{array} 
\right.
\end{equation}
When $x^* = 0$, $g_1(x)$ and $g_2(x)$ do not intersect (see, for example, panel (c) in Figure \ref{fig:prop}). In all other cases, $x^*$ is their intersection point, i.e., the point at which the proportion of senescent and non-senescent deaths is equal. 

\vskip 1cm



\section{Applications} \label{Applications} 
Expressing Makeham models (\ref{hazard-general}) as a mixture can be advantageous for analyzing the components of mortality, e.g., the ones capturing premature and senescent deaths. To illustrate these advantages, we estimate the Gompertz-Makeham model using raw death counts and exposures after age 20 from the Human Mortality Database \citep{hmd} for France, Italy, Japan, and Sweden, years 1947 to 2020, males and females separately.

\cite{mazzuco2021measure} highlight that the share of premature deaths is also defined by the shape, scale, and location of the senescent deaths distribution. This suggests a recommendation to maintain consistent senescent mortality parameters across countries while allowing for variability in premature mortality parameters. However, we will not keep the senescent mortality parameters constant for four main reasons: (i) the population in two different years is comprised of individuals from different cohorts with diverse life histories; (ii) distinct genetic backgrounds, shaped by historical migrations and evolutionary processes, introduce variations in susceptibility to age-related diseases, potentially influencing senescent mortality; (iii) the interplay between genetic factors and environmental determinants, including diet, lifestyle, pollution, and healthcare access, amplifies differences in senescent mortality patterns among countries; and (iv) disparities in healthcare systems and socioeconomic factors add complexity, influencing disease prevention, diagnosis, and treatment effectiveness among countries.

To estimate the parameters, we apply a Bayesian procedure \citep{gelman2013bayesian} and assume a Poisson distribution for the death counts $D_{ijk}$, where the multi-index $ijk$ represents age $i$ in year $j$ for country $k$ \citep[see, for example, ][]{brillinger1986biometrics}. The Bayesian estimates are obtained by the mode of the posterior distributions, also known as the maximum a posteriori probability (MAP) estimate \citep{patricio2023using}. 

The prior (and hyper-prior) distributions are defined as 
\begin{eqnarray*}
    a_{ijk}|\alpha_1, \beta_1 &\sim& InverseGamma(\alpha_1, \beta_1);\\
    b_{ijk}|\alpha_2, \beta_2 &\sim& InverseGamma(\alpha_2, \beta_2);\\
    c_{ijk}|\alpha_3, \beta_3 &\sim& InverseGamma(\alpha_3, \beta_3);\\
    \alpha_1 &\sim& Gamma(2,2);\\
    \alpha_2 &\sim& Gamma(2,2);\\
    \alpha_3 &\sim& Gamma(2,2);\\
    \beta_1 &\sim& Gamma(1,1);\\
    \beta_2 &\sim& Gamma(1,1);\\
    \beta_3 &\sim& Gamma(1,1) \,,
\end{eqnarray*}
where $a$, $b$ and $c$ are the Gompertz-Makeham model parameters. Since the parameters are strictly positive, we choose an inverse-gamma prior distribution. The latter is characterized by a heavy tail and keeps probability further from zero than the Gamma distribution. 

By specifying the prior distributions, we employed the \texttt{NUTS} algorithm, a variant of Hamiltonian Monte Carlo known as the No-U-Turn Sampler \citep{hoffman2014no, betancourt2017conceptual}, via the \texttt{Rstan} R-package \citep{stan} to sample from the posterior distributions. Four chains were executed, each consisting of 6000 iterations (4000 warm-ups and 2000 sampling). We assessed convergence using the R-hat diagnostic and found that none of the R-hat values exceeded 1.05 \citep[see, for instance,][]{vehtari2021rank}. We opted not to thin the chains, as thinning is typically unnecessary with the efficient Hamiltonian Monte Carlo method employed by Stan. The point estimates provided are MAP estimates, and the intervals displayed represent 95\% Highest Density Intervals (HPD intervals) for the posterior distributions.

\begin{figure}[htb!]
    \centering
    \includegraphics[width=\textwidth]{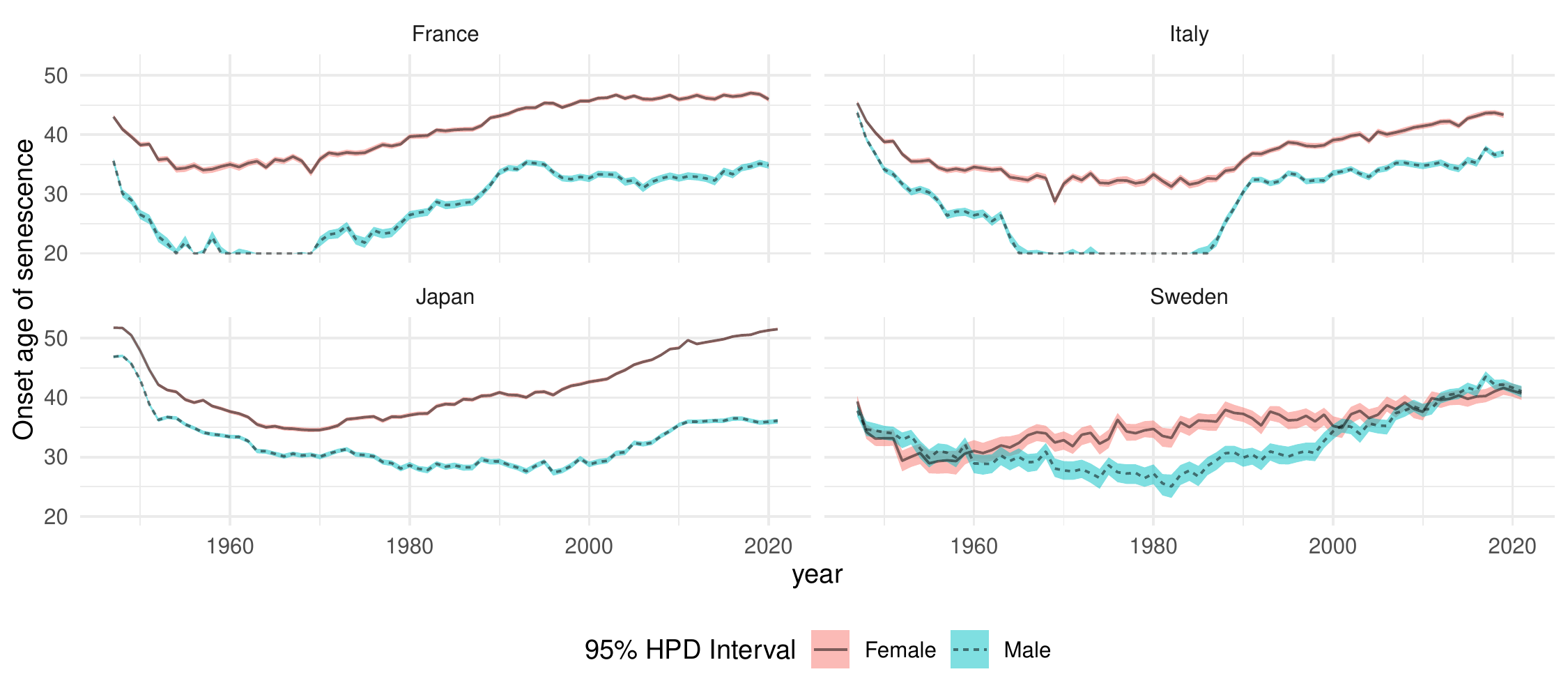}
    \caption{Threshold age between premature and senescence mortality estimated through the Gompertz-Makeham model}
    \label{fig:threshold_estim}
\end{figure}

Figure \ref{fig:threshold_estim} presents the threshold age of senescent mortality within the Gompertz-Makeham framework. In France, post-2000, the threshold age remains relatively stable, fluctuating between approximately 30 and 35 years for males and around 45 years for females. Almost the same holds for Japanese males after 2010. For other populations, notably after the 1990s, the threshold age exhibits a consistent linear increase. This trend suggests that in recent years, senescent mortality prevalence has been postponed at an approximate rate of 3 months per year for both Italian males and females, French males, and Japanese females. Swedish females experience a delay of around 2 months per year, while Swedish males see a more significant postponement of approximately 4 months per year. Japanese males and French females show a slower pace of postponement, at about half a month per year.

{}

\begin{figure}[htb!]
    \centering
    \includegraphics[width=\textwidth]{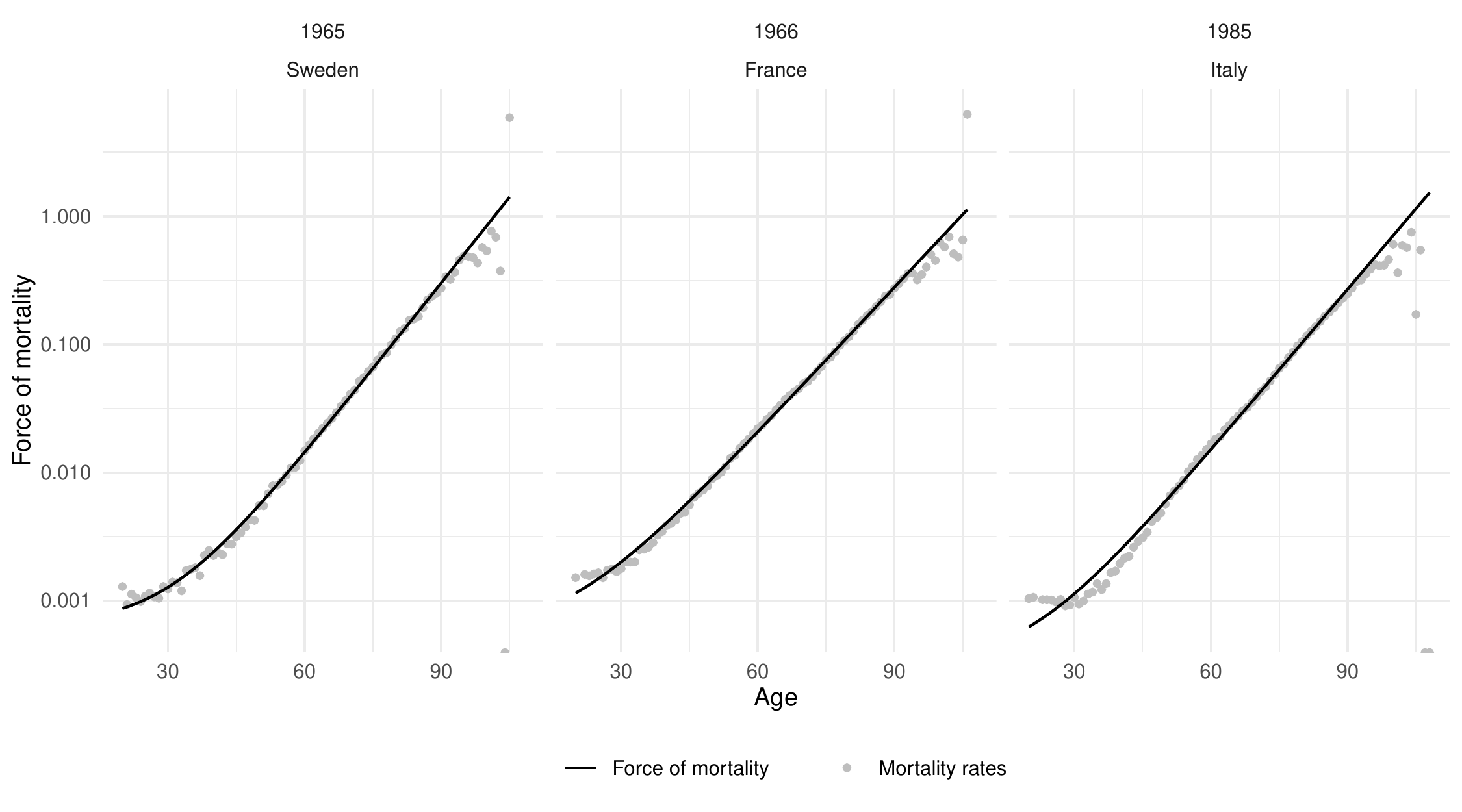}
    \caption{Fitted versus observed mortality rates for males from Sweden in 1965, France in 1966, and Italy in 1985.}
    \label{fig:fit_model}
\end{figure}

In Figure \ref{fig:threshold_estim}, we observe a consistently low threshold age for males in France (from 1960 to 1998) and Italy (from 1965 to 1985). During these periods, the data show a complexity that the Gompertz-Makeham model, shown in Figure \ref{fig:fit_model}, fails to capture. While the Gompertz-Makeham model is simplistic and fits well the log-linear increase in the risk of dying, it struggles to accommodate non-monotonic deviations from this pattern in the starting ages of analysis \citep{MisNem15}. This limitation leads to an underestimation of the extrinsic risk parameter $c$. Consequently, while the Gompertz-Makeham model adequately represents the data pattern for Sweden, it notably fails to provide an accurate fit for France and Italy. This discrepancy is primarily attributed to the complex data pattern observed between ages 20 and 45, which deviates from both the log-linear increase in the risk of dying and the assumption of constant premature mortality risk. As a result, the estimated threshold age $x^*$ is simply the starting age of analysis (see Figure \ref{fig:threshold_estim}).


{}

\begin{figure}[htb!]
    \centering
    \includegraphics[width=\textwidth]{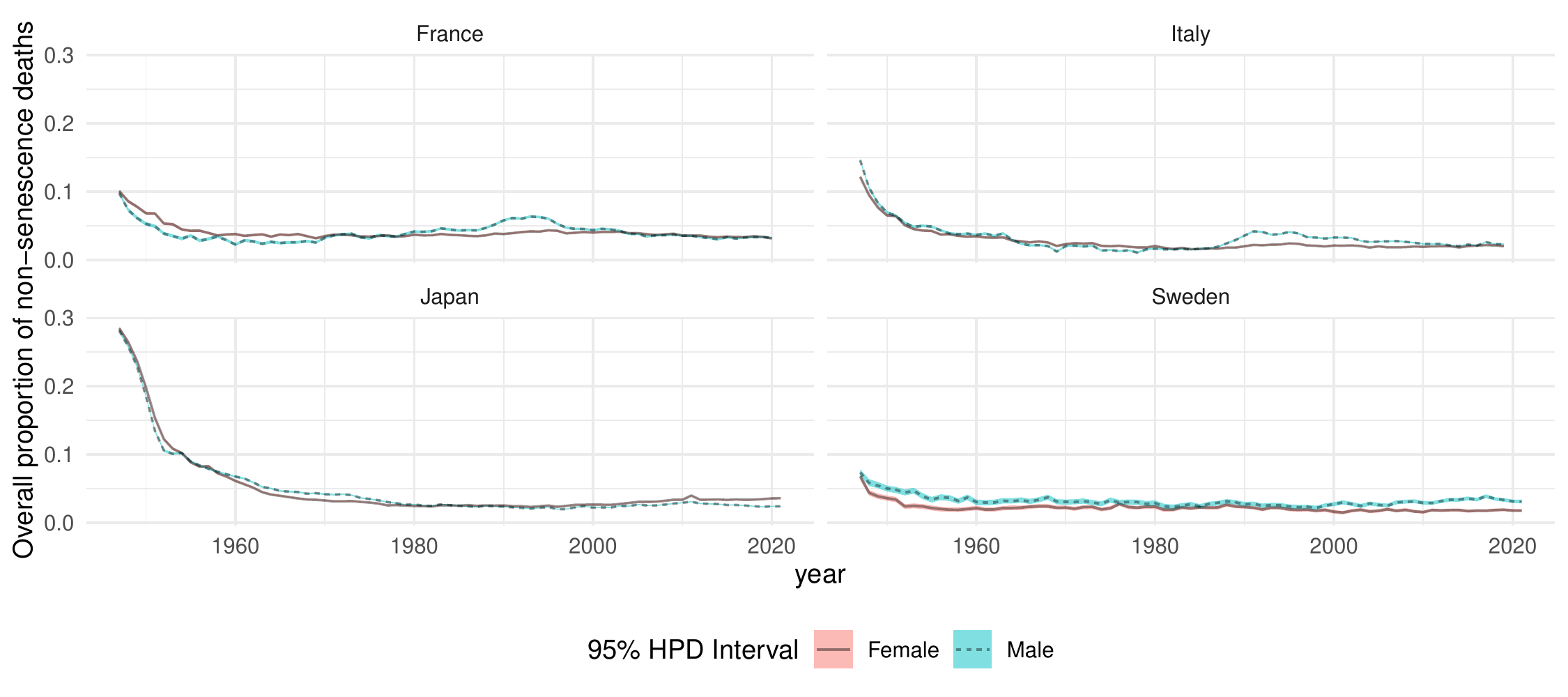}
    \caption{Overall prevalence of premature mortality after age 20 estimated through the Gompertz-Makeham model}
    \label{fig:prev_estim}
\end{figure}

The prevalence of premature mortality after age 20 decreases over time (see Figure \ref{fig:prev_estim}), faster until about 1970 and more moderately afterward, reaching an almost constant level in recent years. Sex-specific premature mortality seems to be balanced, with Sweden having a prevalence for males slightly higher than the one for females over the entire period. For French and Italian males, we observe a sudden increase from 1985 to 1990, leading to a persistent gap between male and female premature mortality prevalence until the late 1990s for France and about a decade later for Italy. This period coincides with the emergence of the global HIV/AIDS epidemic, impacting, in particular, France and Italy \citep[see, for example][]{hamers1998diversity}. During this period, both countries experience a substantial rise in HIV/AIDS-related deaths. Meanwhile, although Sweden and Japan were also affected by the epidemic, the prevalence of HIV/AIDS was much lower compared to Italy and France \citep{gibney2006preventing}. Although advancements in HIV/AIDS treatment emerged, access to these treatments varied. It was not until the later years of the 1990s that antiretroviral therapy became more widely accessible in France and Italy \citep{piot2007hiv, ippolito2001changing}.

{}

\begin{figure}[htb!]
    \centering
    \includegraphics[width=\textwidth]{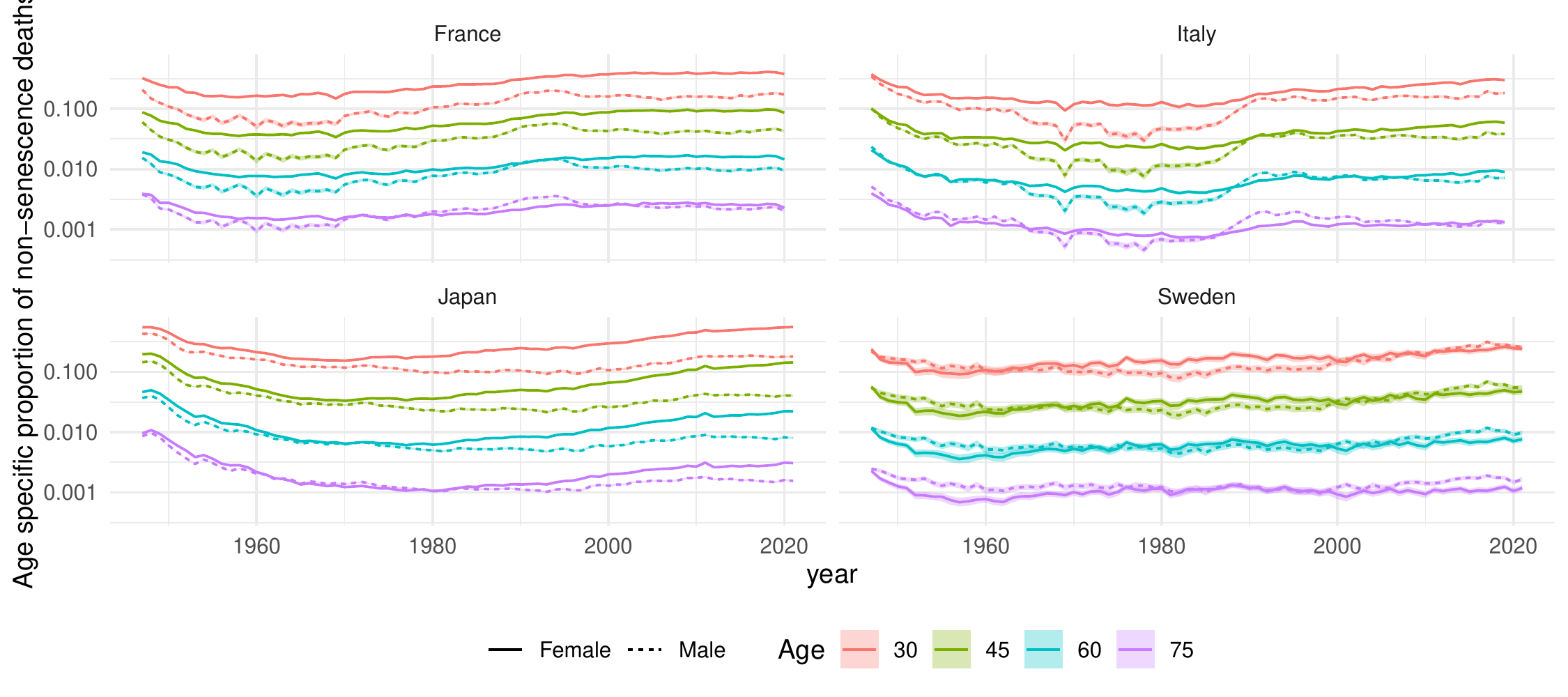}
    \caption{Age-specific prevalence of premature mortality estimated through the Gompertz-Makeham model}
    \label{fig:as_prev_estim}
\end{figure}

Despite the overall prevalence of premature mortality converging for some countries and diverging for others, when we look at the age-specific prevalence of premature mortality (Figure \ref{fig:as_prev_estim}), we see a different picture. As expected, the prevalence of premature mortality decreases with age. However, for France, while the inter-sex difference in overall prevalence seems to converge to zero, we do not observe this for the age-specific prevalence at younger ages. The same trend is seen in Italy and Japan. For Sweden, however, what we see is the opposite. The prevalence seems to be the same for males and females between ages 30 and 45, and slightly higher for males from ages 60 to 75. 

{}

Over time, the age-specific prevalence of premature mortality for the French population seems to be stable for both sexes after 2000. The same holds for Japanese males after 2010 and for Italian males after 1990. For Sweden, the trend in age-specific prevalence of premature mortality seems to be identical for both sexes. Other mortality measures, such as the premature and senescent distributions of deaths, their force of mortality at ages 40, 60, and 80, and their modal age at death, are presented in the appendix.

\section{Conclusion}


In this paper, we present a formal relationship that represents competing risk models as a mixture of distributions, where each cause of death corresponds to a component distribution, and an individual's lifespan is a mixture of these component distributions. The weights in this mixture correspond to the probabilities of each cause of death being the first to occur among all competing causes. This representation allows for a clear understanding of how competing risks influence overall mortality and provides a framework for modeling complex mortality patterns.

The mixture-model specification aids in representing the distribution of deaths as a convex combination of distributions for risk-specific subpopulations. This facilitates calculating various mortality and longevity measures for each subpopulation, as well as assessing the overall and age-specific prevalence of each cause of death. 

We focus on the special case of Makeham mortality models that incorporate a constant extrinsic risk of dying. We represent these models as a mixture that reflects individual lifetimes in a competing-risk setting. The interpretation is straightforward: an individual dies either according to a baseline mortality mechanism or an exponential distribution. 

In the case of a two-risk model, the Gompertz-Makeham, when components capture senescent and premature deaths, we can estimate the threshold age that marks the change of death prevalence from premature to senescent. We are also able to reconstruct the age-specific profiles of the premature and senescent mortality components. To illustrate these findings, we take advantage of a Bayesian approach to estimate the Gompertz-Makeham model for the French, Italian, Japanese, and Swedish populations from 1947 to 2020, after age 20. The results suggest a postponement of senescent mortality prevalence at a pace between 2 and 4 months per year for most of the populations included in our analysis. The overall prevalence of premature mortality differs across age groups and between sexes. 

\section{Acknowledgements}
The research leading to this publication is part of a project that has received funding from the European Research Council (ERC) under the European Union’s Horizon 2020 research and innovation program (Grant agreement No. 884328 – Unequal Lifespans).  Silvio C Patricio gratefully acknowledges the support from the AXA Research Fund through funding the "AXA Chair in Longevity Research."


\newpage
\bibliography{mybib}
\bibliographystyle{apalike}

\nocite{*}


\clearpage
\section*{Appendix} 
\addcontentsline{toc}{section}{\numberline{}Appendix} 
\setcounter{section}{0}
\renewcommand{\thesection}{A-\arabic{section}}
\setcounter{figure}{0}
\renewcommand{\thefigure}{A-\arabic{figure}}
\setcounter{table}{0}
\renewcommand{\thetable}{A-\arabic{table}}

\begin{figure}[htb!]
    \centering
    \includegraphics[width=\textwidth]{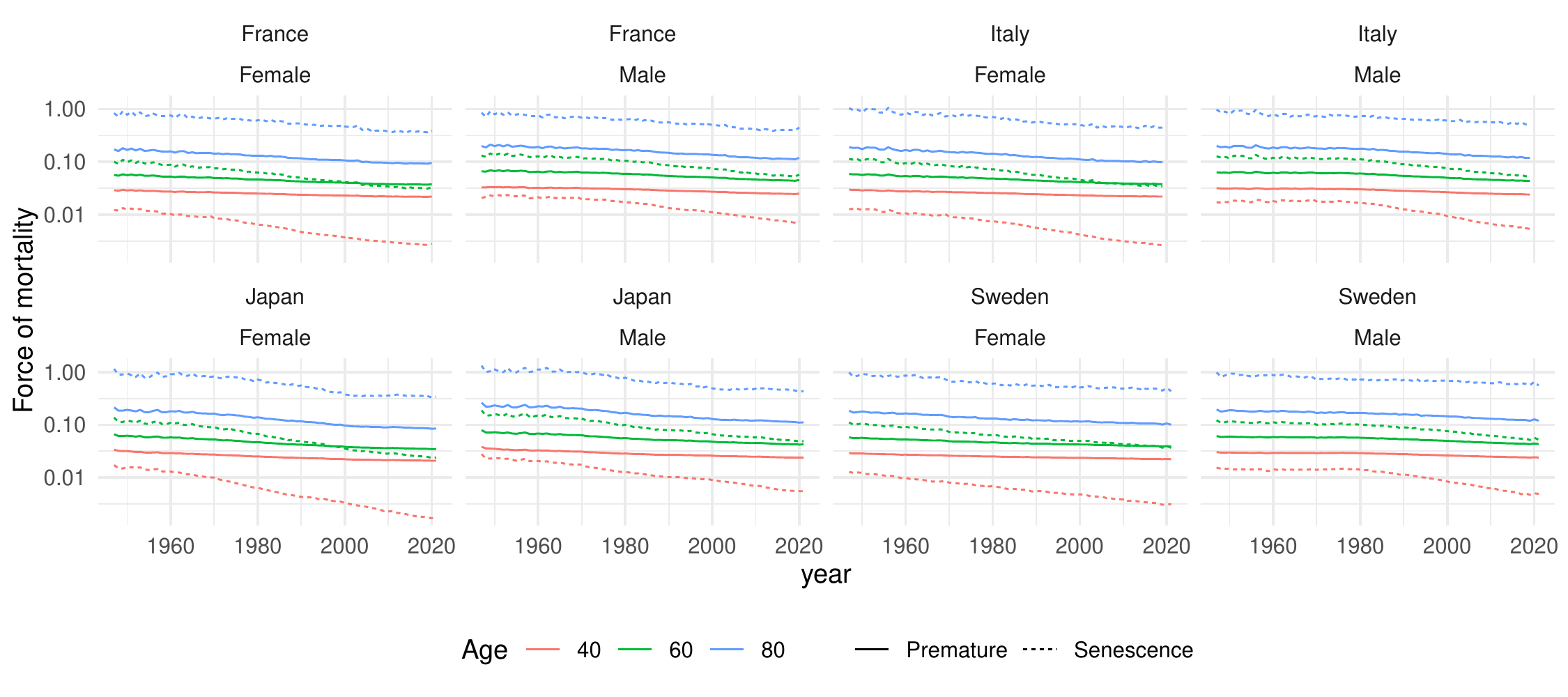}
    \caption{Estimated force of mortality at ages 40, 60, and 80 through the Gompertz-Makeham model for premature and senescent groups}
    \label{}
\end{figure}{}

\begin{figure}[htb!]
    \centering
    \includegraphics[width=\textwidth]{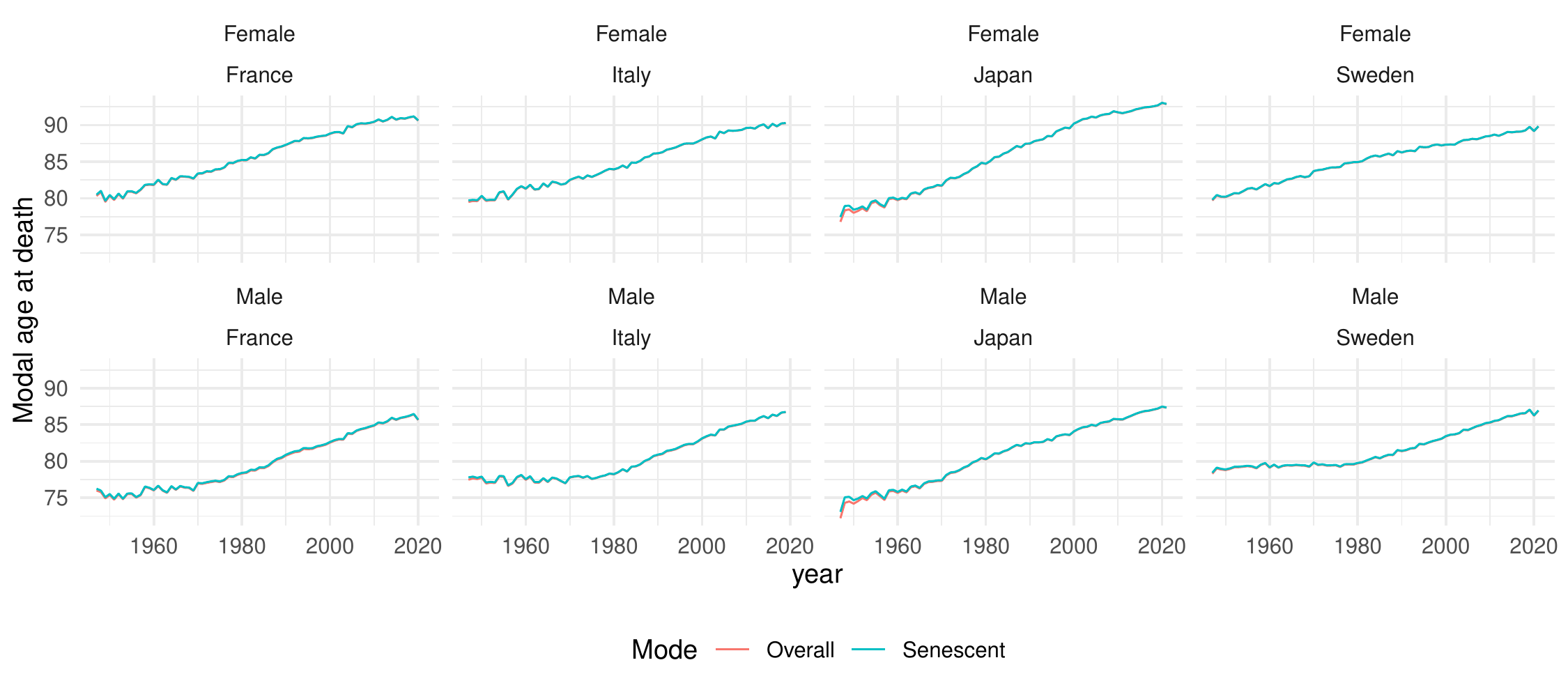}
    \caption{Estimated senescence and overall modal age at death through the Gompertz-Makeham model}
    \label{}
\end{figure}{}

\begin{figure}[htb!]
    \centering
    \includegraphics[width=\textwidth]{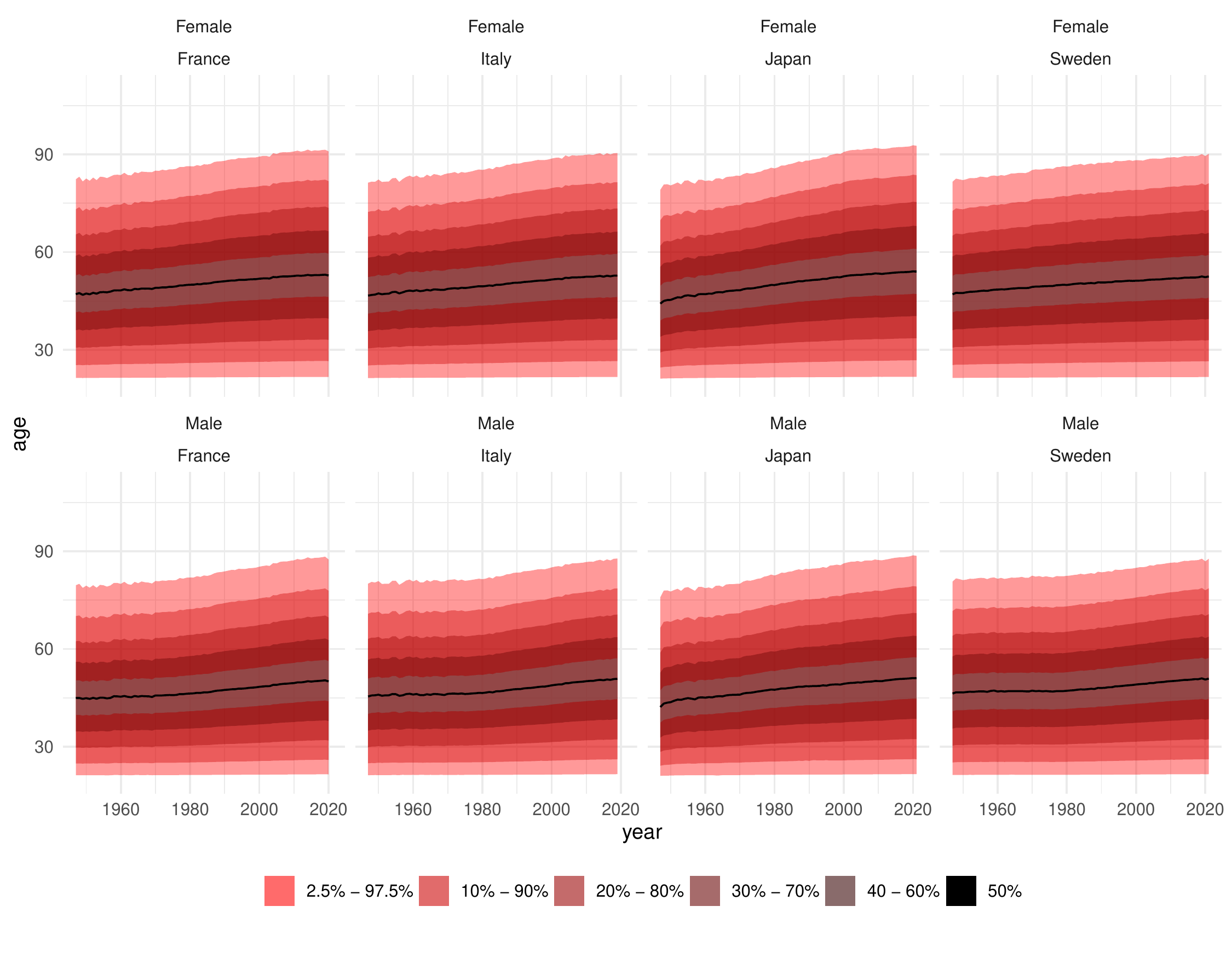}
    \caption{Estimated non-senescent death distribution after age 20 through the Gompertz-Makeham model}
    \label{}
\end{figure}{}

\begin{figure}[htb!]
    \centering
    \includegraphics[width=\textwidth]{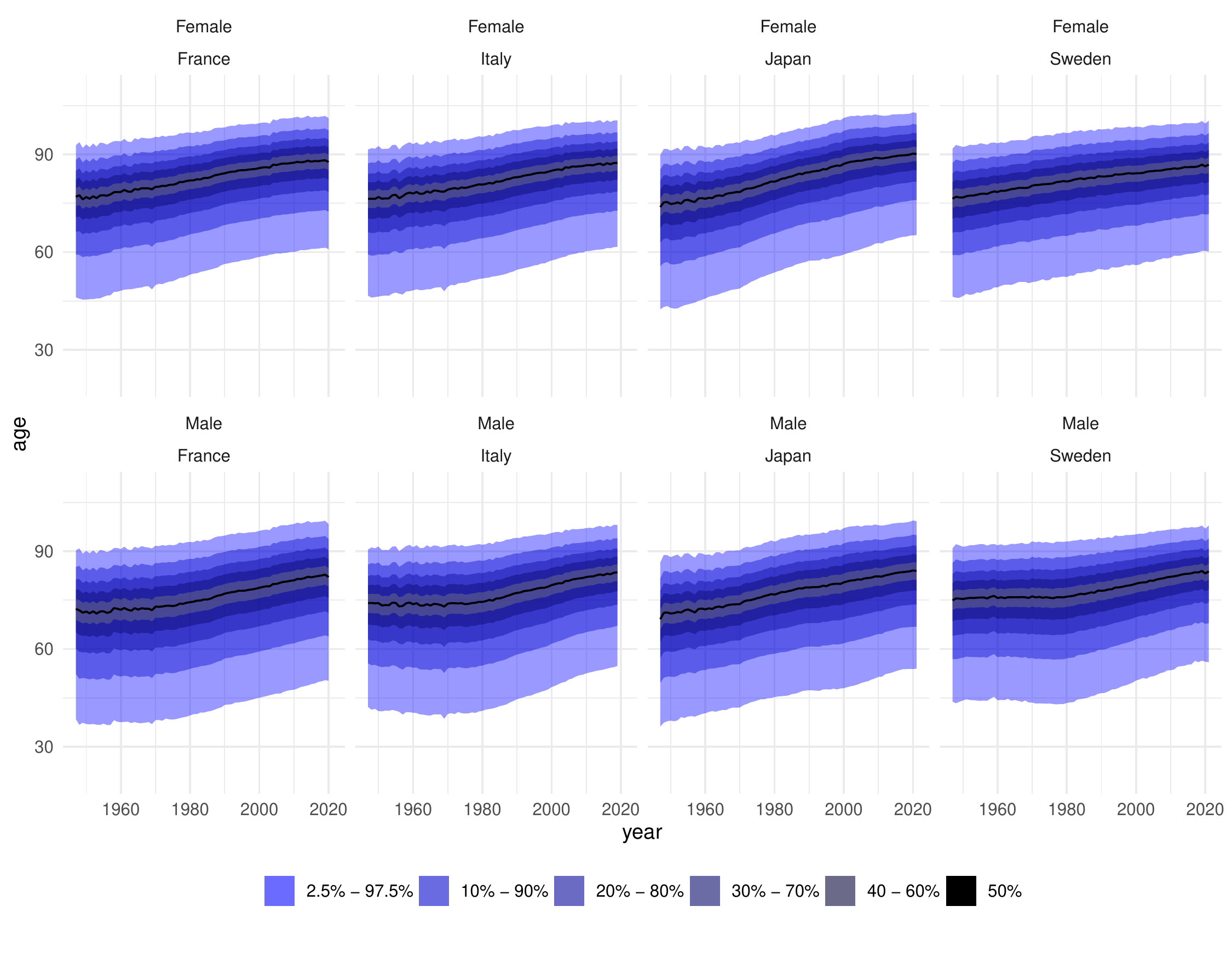}
    \caption{Estimated non-senescent death distribution through the Gompertz-Makeham model}
    \label{}
\end{figure}{}





\cleardoublepage
\end{document}